%Paper: hep-th/9509070
%From: Clifford Johnson <cvj@itp.ucsb.edu>
%Date: Wed, 13 Sep 95 18:01:06 PDT
%Date (revised): Mon, 30 Oct 95 19:45:12 PST

%%%%%%%%%%%%%%% cut here %%%%%%%%%%%%%%%%%%%%%%%%%%%%%
\input harvmac

%%%%%%%%%%%%%%%%%%
\Title{\vbox{\baselineskip12pt
\hbox{NSF-ITP-95-108}
\hbox{McGill/95-45}
\hbox{CERN-TH/95-236}
\hbox{PUPT-1561}
\hbox{hep-th/9509070}
}}
{\vbox{\centerline{Is string theory a theory of strings?}}}

\centerline{Clifford V. Johnson$^1$,
Nemanja Kaloper$^2$,
Ramzi R.~Khuri$^{2,3}$ and Robert C.~Myers$^2$}
\bigskip\centerline{$^1${\it Institute for Theoretical Physics, UCSB,
CA~93106, USA }}
\bigskip\centerline{$^2${\it Physics Department, McGill
University, Montreal, PQ, H3A 2T8 Canada}}
\bigskip\centerline{$^3${\it Theory Division, CERN,
 CH-1211, Geneva 23, Switzerland}}
\vskip .3in
Recently a great deal of evidence has been found indicating
that type $IIA$ string theory compactified on $K3$
is equivalent to heterotic string theory compactified on $T^4$.
Under the transformation which relates
the two theories, the roles of fundamental and solitonic
string solutions are interchanged.
In this letter we show that there exists a solitonic membrane
solution of the heterotic string theory which becomes a singular
solution of the type $IIA$ theory, and
should therefore be interpreted as a
fundamental membrane in the latter theory. We speculate
upon the implications that the complete type $IIA$ theory is
a theory of membranes, as well as strings.

\vskip .3in
\Date{\vbox{\baselineskip12pt
\hbox{September 1995 (Revised Version)}}}

\def\sqr#1#2{{\vbox{\hrule height.#2pt\hbox{\vrule width
.#2pt height#1pt \kern#1pt\vrule width.#2pt}\hrule height.#2pt}}}

\def\a{\alpha}

\def\ie{{\it i.e.,}\ }
\def\eg{{\it e.g.,}\ }

\def\[{\bf (*}
\def\]{*) \rm\ }

\lref\nelson{W.~Nelson, Phys. Rev. {\bf D49} (1994) 5302
[hep-th/9312058].}

\lref\stst{M.~J.~Duff and R.~R.~Khuri, Nucl. Phys. {\bf B411}
(1994) 473 [hep-th/9305142]; M.~J.~Duff, Nucl. Phys. {\bf B442}
(1995) 47 [hep-th/9501030].}

\lref\hetii{C.~M.~Hull and P.~K.~Townsend, Nucl. Phys. {\bf B438}
(1995) 109 [hep-th/9410167].}

\lref\ed{E.~Witten, Nucl. Phys. {\bf B443}
(1995) 85 [hep-th/9503124].}

\lref\probes{M.~J.~Duff, R.~R.~Khuri and J.~X.~Lu,
Nucl. Phys. {\bf B377} (1992) 281 [hep-th/9112023].}

\lref\prep{M.~J.~Duff, R.~R.~Khuri and J.~X.~Lu,
Phys. Rep. {\bf 259} (1995) 213 [hep-th/9412184].}

\lref\ccount{A.~Achucarro, J.~Evans, P.K.~Townsend and D.~Wiltshire,
Phys. Lett. {\bf B198} (1987) 441.}

\lref\becker{K.~Becker, M.~Becker and A.~Strominger,
``Fivebranes, membranes and nonperturbative string theory'',
preprint [hep-th/9507158].}

\lref\tseyt{A.A. Tseytlin, ``On singularities of spherically symmetric
backgrounds in string theory", preprint [hep-th/9509050].}

\lref\hmono{R.~R.~Khuri, Phys. Lett. {\bf B259} (1991) 261;
Nucl. Phys. {\bf B387} (1992) 315 [hep-th/9205081];
M.~J.~Duff, R.~R.~Khuri,
R.~Minasian and J. Rahmfeld, Nucl. Phys. {\bf B418} (1994) 195
[hep-th/9311120].}

\lref\bhep{M.~J.~Duff and J. Rahmfeld, Phys. Lett.
{\bf B345} (1995) 441 [hep-th/9406105]; R.~R.~Khuri and R.~C.~Myers,
``Dynamics of extreme black holes and massive string states'',
preprint [hep-th/9508045].}

\lref\town{P.K.~Townsend, Phys. Lett. {\bf B350} (1995) 184
[hep-th/9501068]; Phys. Lett. {\bf B354} (1995) 247 [hep-th/9504095];
C.M.~Hull and P.K.~Townsend, ``Enhanced gauge symmetries in superstring
theories'', preprint [hep-th/9505073].}

\lref\towns{P.~K.~Townsend,
 ``$P$-Brane Democracy'', preprint [hep-th/9507048].}

\lref\schsen{J.~H.~Schwarz and A.~Sen, Phys. Lett. {\bf B312}
(1993) 105 [hep-th/9305185]; A.~Sen, Int. J. Mod. Phys.
{\bf A9} (1994) 3707 [hep-th/9402002].}

\lref\ghs{D.~Garfinkle, G.~T.~Horowitz and A.~Strominger,
Phys. Rev. {\bf D43} (1991) 3140; erratum {\bf D45} (1992) 3888.}

\lref\senb{A.~Sen, Nucl. Phys. {\bf B440} (1995) 421
[hep-th/9411187].}

\lref\gibbons{G.~W.~Gibbons, Nucl. Phys. {\bf B207} (1982) 337.}

\lref\solitons{A.~Sen, ``String string duality conjecture in
six dimensions and charged solitonic strings'',
preprint [hep-th/9504027]; J.~A.~Harvey and A.~Strominger,
``The Heterotic String is a Soliton'', preprint
[hep-th/9504047].}

\lref\seiberg{N.~Seiberg and E.~Witten, Nucl. Phys. {\bf B426} (1994)
19; erratum, {\bf B430} (1994) 485 [hep-th/9407087];
Nucl. Phys. {\bf B431} (1994) 484 [hep-th/9408099].}

\lref\five{M.~J.~Duff and J.~X.~Lu, Nucl. Phys. {\bf B354}
(1991) 141.}

\lref\throat{C.~G.~Callan, J.~A.~Harvey and A.~Strominger,
Nucl. Phys. {\bf B359} (1991) 611; in Proceedings of {\it
String Theory and Quantum Gravity '91} (Trieste) 208-244
[hep-th/9112030].}

\lref\strom{A.~Strominger, Nucl. Phys. {\bf B343} (1990) 167.}

\lref\klopv{R.~Kallosh, A.~Linde, T.~Ortin, A.~Peet and
A.~Van Proeyen, Phys. Rev. {\bf D46} (1992) 5278
[hep-th/9205027].}

\lref\curtright{T.~L.~Curtright and C.~K.~Zachos,
Phys. Rev. Lett. {\bf 53} (1984) 1799;
E.~Braaten, T.~L.~Curtright and C.~K.~Zachos,
Nucl. Phys. {\bf B260} (1985) 630.}

\lref\linear{R.~C.~Myers, Phys. Lett. {\bf B199} (1987) 371;
S.~P.~de Alwis, J.~Polchinski and R.~Schimmrigk,
Phys. Lett. {\bf B218} (1989) 449; I.~Antoniadis, C.~Bachas,
J.~Ellis and D.~V.~Nanopoulos, Phys. Lett. {\bf 211B} (1988)
393.}

\lref\rey{S.~J.~Rey, Phys. Rev. {\bf D43} (1991) 526.}

\lref\ortin{R.~Kallosh and T.~Ortin, Phys. Rev. {\bf D50} (1994) 7123
[hep-th/9409060].}

\lref\peet{A.Sen, ``Extremal black holes and elementary string
states'', preprint [hep-th/9504147];
A.~Peet, ``Entropy and supersymmetry of $D$
dimensional extremal electric black holes versus string states'',
preprint [hep-th/9506200]; J.~Russo and L.~Susskind, ``Asymptotic
level density in heterotic string theory and rotating black
holes'', preprint [hep-th/9405117].}

\lref\mont{C.~Montonen and D.~Olive, Phys. Lett. {\bf 72B}
(1977) 117.}

\lref\poly{A.~M.~Polyakov, Phys. Lett. {\bf 103B} (1981) 207,
211.}

\lref\stfb{M.~J.~Duff, Class. Quant. Grav. {\bf 5} (1988) 189;
M.~J.~Duff and J.~X.~Lu, Nucl. Phys. {\bf B354} (1991) 129.}

\lref\nonren{P.~S.~Howe and G.~Papadopoulos, Nucl. Phys.
{\bf B289} (1987) 264; Class Quant. Grav. {\bf 5} (1988) 1647.}

\lref\tension{A.~Dabholkar and J.~A.~Harvey, Phys. Rev. Lett.
{\bf 63} (1989) 478; A.~Dabholkar, G.~Gibbons, J.~A.~Harvey and F.~Ruiz
Ruiz, Nucl. Phys. {\bf B340} (1990) 33.}

\lref\enhanced{C.~M.~Hull and P.~K.~Townsend, ``Enhanced Gauge
Symmetries
in superstring Theories'', preprint [hep-th/9505073];
P.~S.~Aspinwall, ``Enhanced Gauge Symmetries and $K3$ Surfaces'',
preprint [hep-th/9507012]; E.~Witten, ``Some Comments on String
Dynamics'', preprint [hep-th/9507121].}

\lref\heti{A.~Dabholkar, ``Ten dimensional heterotic string as a
soliton'', preprint [hep-th/9506160]; C.~M.~Hull, ``String-string
duality in ten dimensions'', preprint [hep-th/9506194].}

\lref\moduli{K.~Narain, Phys. Lett. {\bf 169B} (1986) 41;
N.~Seiberg, Nucl. Phys. {\bf B303} (1988) 286;
P.S.~Aspinwall and D.R.~Morrison, ``String Theory on K3 Surfaces,''
[hep-th/9404151].}

\lref\vafa{C.~Vafa and E.~Witten, Nucl. Phys. {\bf B447} (1995) 261
[hep-th/9505053].}

\lref\sharp{E.~R.~Sharpe and C.~V.~Johnson, unpublished.}

\lref\cone{A.~Strominger, ``Massless black holes and conifolds
in string theory,'' preprint [hep-th/9504090].}

\lref\convent{J.~Hughes, J.~Liu and J.~Polchinski, Phys. Lett.
{\bf B180} (1986) 370; E.~Bergshoeff, E.~Sezgin and P.K.~Townsend,
Phys. Lett. {\bf B189} (1987) 75.}

\lref\dpage{D.N.~Page, Phys. Lett. {\bf B80} (1978) 55.}

%%%%%%%%%%%%%%%%%%%%%%%%%%%%%%%%%%%%%%%%%%%%%%%%%%%%%%%%%%%%%%%%

Beyond the first quantized framework of the Polyakov path
integral \poly, our knowledge of string theory is sorely
lacking. Determining a more fundamental formulation of
the theory will be an essential step in fully addressing
nonperturbative issues, such as supersymmetry breaking
and selection of the vacuum. Recently there has been
rapid progress into understanding the strong coupling
dynamics of certain supersymmetric string theories \ed.
These advances
should provide new insights into the correct
fundamental framework which must underly string theory.
We take one step in this direction by presenting evidence that type
$IIA$ superstrings are only one component of a larger
theory which also contains fundamental membranes.

The recent developments seem to indicate that the strong
coupling physics of certain superstring
theories may be reformulated as the weak coupling
physics of ``dual'' string
theories \refs{\ed,\stst,\hetii}
and {\it vice versa}.  One interesting example
is the duality in six dimensions between heterotic strings
compactified
on a four-torus $T^4$, and type $IIA$ superstrings compactified on
a particular four-manifold known as $K3$.
In fact, the duality relies on the much stronger conjecture that
these two string are completely equivalent. Some of the evidence
supporting this conjecture is: (i) both theories have the same
supersymmetric multiplets of massless states \refs{\hetii,\ed};
(ii) the moduli space
of the possible vacua is identical in both theories \moduli; (iii)
the low energy effective actions can be identified with
a strong-to-weak coupling duality mapping of the massless fields
\ed;
(iv) there is one-loop consistency via certain anomaly constraints
\vafa.

One further result which supports the equivalence
is that the heterotic string can be identified
as a soliton within the type $IIA$ string theory, and conversely,
the type $IIA$ string can be identified
as a soliton in heterotic string theory\foot{This second identification
is
made only at the level of external  field configurations, in
the first of the references in \solitons.}\ \solitons.
Thus under the duality
transformation, the roles of the fundamental and solitonic strings
are interchanged. This interchange is a stringy version of the
role reversal between magnetic monopoles and electric charges
arising in the strong-to-weak coupling duality of
gauge field theories conjectured in \mont, and recently
confirmed in the context of supersymmetric field theory \seiberg.

In general, the low energy string theories contain a rich
array of solutions corresponding to extended
objects, so-called $p$-branes for $p$-dimensional bodies
(see \prep\ and references therein).
It is now apparent that
these objects play an important role in the non-perturbative
physics of the string theories \becker. One is then prompted to ask
how these solutions behave under the strong/weak coupling
duality transformations discussed above.
There will be three distinct possibilities: (i) the $p$-brane could be
a singular field configuration in both of the dual string theories,
which would justify discarding these configurations
as unphysical, (ii) the $p$-brane could be nonsingular in both theories,
in which case it would be treated as a soliton in both contexts,
and finally (iii) the $p$-brane could be nonsingular in one
theory but singular in the dual theory. In the latter case, since
it appears as a soliton in one theory, one would not be able to
omit it from the spectrum. However the fact that the $p$-brane
solution is singular in the dual theory suggests that it represents
the external fields around a fundamental
source\foot{In case (i), one could
also consider the possibility that the $p$-brane
is fundamental in both of the theories.} -- \ie the dual
theory should contain fundamental $p$-branes!

An immediate question is how the
singular or nonsingular nature of these objects is to be
determined\foot{Ref. \tseyt\ recently presented a complementary
discussion of singular solutions which stresses the importance
of the source terms. It also conjectures that in the presence of
such sources the singularities may actually be smoothed out with
certain field redefinitions.}.
The result can be phrased in terms of
examining the $p$-brane with a certain test-probe,
\ie determining the behavior of a small test object as it approaches
the core of the $p$-brane. The choice of the test-probe would depend
on which fundamental theory underlies the original brane solution
(see \refs{\probes,\prep}). For example, in a
heterotic string theory, the natural test-probe to examine
any $p$-brane solution would be a fundamental heterotic
string. This amounts to measuring possible curvature
singularities with the metric which couples to the world-volume
of the fundamental objects in the theory, \ie the
metric which appears in the sigma-model describing these
fundamental objects.
Applied to the case of the six-dimensional string/string duality, this
means that the heterotic string appears
singular in the heterotic string sigma-model
metric, but is nonsingular in the type $IIA$ superstring metric
\solitons.

In this letter as a first step we construct a certain
membrane soliton  solution for $D=6$ heterotic string theory.
We choose this solution to be spacetime
supersymmetric, and show that it is nonsingular
to all orders in the $\alpha'$ expansion.
However, the membrane solution is singular
for type $IIA$ string test-probes. Hence it
falls into class (iii), wherefore
we are led to
conclude that the  full type $IIA$ theory includes
fundamental membranes, as well as strings. Finally, we
discuss the possible implications of this result.

We begin by considering the heterotic string theory
arising from toroidal compactification down to six
dimensions. For a generic point in the moduli space, the low
energy effective theory is $N=2$ supergravity coupled to
twenty abelian vector multiplets.
Thus the bosonic fields include the metric, the dilaton, the
antisymmetric Kalb-Ramond field, $24$ abelian gauge fields, and
$80$ scalar moduli fields. Given these fields, it is
straightforward to list the objects which arise naturally
as solutions in the low energy theory.
For example, in six dimensions the three-form field strength of the
Kalb-Ramond two-form couples naturally as the ``electric'' or
``magnetic'' field around a one-brane, or string.
In fact these correspond to the two string solutions discussed above,
\ie the fundamental heterotic string with the electric
Kalb-Ramond charge, and its dual solitonic string, with
the magnetic three-form charge. Point-like or zero-brane solutions
also appear, with conventional electric charges from the $U(1)$
two-form field strengths.
In particular, singular point-like objects arise
as the extremal limits of electrically charged black holes
\peet. In this case, the dual
objects are two-branes or membranes with magnetic $U(1)$ charge.
To complete the list, one could also consider three-branes
which carry a ``magnetic'' charge from the periodic moduli scalars,
and ``minus-one''-branes or instantons carrying scalar electric charge.
We will restrict our attention, though, to a class of solitonic
membranes.

For the solutions which we wish to consider, it is consistent
to truncate the low energy action as follows:
\eqn\hetact{S_{het}=\int d^6\!x\, \sqrt{-G}\, e^{-2\Phi}
\left(R + 4(\partial\Phi)^2 - {1\over 4}F^2\right),}
where $F=dA$ is the field strength for one of
the $U(1)$ gauge fields, $\Phi$ is the six-dimensional dilaton
and
the metric is that which couples to the heterotic string
sigma-model.
For this action, one finds the following solution
which represents a magnetically charged membrane
\eqn\hetmem{\eqalign{ds^2&=-dt^2+dx_1^2+dx_2^2
+\left(1+{Q\over y}\right)^2\left(dy^2+y^2d\Omega_2^2\right),\cr
e^{2\Phi}&=1+{Q\over y}, \cr
F_{\theta\varphi}&=\sqrt{2}Q\sin\theta.\cr}}
Here $(y,\theta,\varphi)$ are polar coordinates on the $(x_3,x_4,x_5)$
subspace,
and $d\Omega_2^2$ is the line element on the unit two-sphere.
Our solution may be more familiar as the magnetically-charged
extreme dilaton black hole from four dimensions
\refs{\ghs,\gibbons},
raised to six dimensions by adding the flat $x_1,x_2$ directions,
which are tangent to the membrane.

While the metric in \hetmem\
may appear singular at the core of the membrane,
this is a coordinate artifact. In fact, the
solution develops an infinitely long throat with a constant
radius
as $y\to 0$, as is most easily recognized with the coordinate
transformation $\rho/Q=\log(y/Q)$. Then the fields near the core
become
\eqn\throatsol{\eqalign{ds^2&\simeq
-dt^2+dx_1^2+dx_2^2+d\rho^2+Q^2\,d\Omega^2_2,\cr
\Phi&\simeq-\rho/2Q, \cr
F_{\theta\varphi}&=\sqrt{2}Q\sin\theta,\cr}}
The above description of the throat geometry is made using
the heterotic string sigma-model metric, and hence this
membrane is completely nonsingular for the heterotic string
test-probes.

We will be interested in considering these solutions in the
strong coupling regime in which the dual type $IIA$ string theory
is weakly coupled. Thus we will seek supersymmetric
membrane solutions saturating a BPS bound, for which
the mass-charge
relations are preserved against higher-order corrections
in the strong coupling regime \tension.
Therefore, while any of the $24$ heterotic gauge fields could be used
in the construction of the solution \hetmem, we restrict our
attention to those constructed with one of the four gauge
fields contained in the supergravity multiplet. This provides
a supersymmetric embedding of \hetmem\ in the full six-dimensional
$N=2$ theory in which half of the spacetime supersymmetries are
preserved \klopv. Again, the
importance of this feature lies in the resultant absence of
quantum corrections due to the existence of nonrenormalization
theorems \tension.
Choosing the gauge field from one of the vector supermultiplets
results in a membrane which breaks all of the spacetime
supersymmetries.

In the context of the ten-dimensional heterotic string theory,
the supersymmetric choice of gauge fields corresponds to setting
$G_{i\mu}=B_{i\mu}=A_\mu$, where $G$ and $B$ denote the ten-dimensional
metric and Kalb-Ramond field, respectively, with $\mu=0,1,2,3,4,5$,
a spacetime index
and $i=6,7,8$ {\it or} 9 corresponding to {\it one} of the directions
compactified on the four-torus. It is interesting to note that with
this embedding the compact $x_i$ direction combines with the
spatial two-sphere to form a Hopf fibration of the three-sphere
\nelson. The ten-dimensional throat solution is then:
a constant radius three-sphere supported by the parallelizing
torsion of the Kalb-Ramond field, a linear dilaton background
in the $\rho$ direction, and five flat spatial directions and
a trivial time direction. This corresponds precisely to the
throat limit \throat\ of the ten-dimensional neutral fivebrane solution
\refs{\five,\throat}, and so
reveals that our membrane is in fact a fivebrane ``warped''
around the toroidally compactified directions\foot{The
solution is a ``warped" as opposed to ``wrapped" fivebrane
\refs{\strom,\stst}. The latter
dimensionally reduces to an $a= \sqrt{3}$
black hole/$H$-monopole \hmono\ in $D=4$
as opposed to the $a=1$ solution we started
with in this paper. In our ``warped'' solution one of the compact
directions is tied up in the three-sphere
surrounding the fivebrane in a topologically nontrivial way.}.

Despite their spacetime supersymmetry, one might expect
that the membrane solutions are modified by corrections
of higher order in the world-sheet $\alpha'$ expansion
since we are working within heterotic string theory.
These modifications could jeopardize the nonsingular
nature of the membrane core,
a property which is central to our discussion.
However, in this case the throat of the membrane core
is described on the heterotic string world-sheet
by an exact
conformal field theory \sharp. In fact, this conformal field
theory corresponds to precisely that which describes the
throat of the symmetric
five-brane \throat, \ie a
supersymmetric $SU(2)$ Wess-Zumino-Witten model together with a
linear dilaton in the radial direction.
Thus the throat solution \throatsol\ is essentially unchanged, and one
is guaranteed that no singularities develop at the membrane core.
Thus despite the appearance of $\a'$ corrections, we are assured
that the membrane is a stable soliton of the heterotic string
theory\foot{Note that for vacua with enhanced gauge symmetry,
the leading order solution \hetmem\ can be elevated to an exact
solution by the addition of an $SU(2)$ Yang-Mills vector and scalar
at order $\a'$ \ortin. In the context of the ten-dimensional theory,
these new fields correspond to the appearance of a non-abelian gauge
field which cancels the gravitational part of the $\a'$ corrections.}.
We also expect that the background Killing spinors are perturbatively
corrected so that spacetime supersymmetry also survives the $\a'$
corrections.
This is certainly the case in the throat region where the exact
conformal field theory description applies.

Now consider transforming the membrane soliton to the type $IIA$
string theory via the ``duality'' mapping indicated in \ed:
\eqn\duality{\Phi'=-\Phi, \qquad\qquad
G_{\mu\nu}'=e^{-2\Phi} G_{\mu\nu},\qquad\qquad
A_\mu'=A_\mu.}
Here the (un)primed fields are those arising in the type $IIA$
(heterotic) string theory. In particular, $G_{\mu\nu}'$
is the metric which couples to the type $IIA$ string sigma-model.
The type $IIA$ action is then given by
\eqn\iiact{S_{IIA}=\int d^6\!x\, \sqrt{-G'} \left[e^{-2\Phi'}
\left(R' + 4(\partial\Phi')^2\right) -
{1\over 4}F'^2\right],}
and the solution becomes
\eqn\iimem{\eqalign{ds'^2&=\left(1+{Q\over y}\right)^{-1}
\left(-dt^2+dx_1^2+dx_2^2\right)
+\left(1+{Q\over y}\right)\left(dy^2+y^2d\Omega_2^2\right),\cr
e^{2\Phi'}&=\left(1+{Q\over y}\right)^{-1}, \cr
F_{\theta\varphi}'&=\sqrt{2}Q\sin\theta.\cr}}
In this frame the leading order solution becomes singular,
requiring sources to support it at the core. First, the core,
{\it i.e.,}\ $y=0$, is  a finite proper distance away, and the curvature
diverges there, {\it e.g.,}\ the Ricci scalar goes as $R\sim 1/(Qy)$.
Thus from the point of view of type $IIA$
string test probes, the membrane appears singular.
Essentially with \duality, we have made a singular conformal
transformation of the original metric which implicitly
adds an extra ``point-at-infinity'' closing off the end of the throat.
To consistently solve the new equations of motion for \iiact, we must
now include a source at this end-point, \ie $y=0$.
Hence in the type $IIA$ theory, the membrane must be interpreted
as fundamental.

In summary,
we began by constructing a nonsingular supersymmetric
solution in the heterotic string theory, which represents a
magnetically charged membrane. Because of the nonsingular nature
of the solution, it appears that these field configurations
must be included in defining the heterotic string
theory. Mapped to the type $IIA$ theory via \duality, these
solutions require a source so
become singular suggesting that they should be
interpreted as fundamental membranes within the type
$IIA$ theory\foot{It may be that the duality
transformation \duality\ is incomplete, and that it is
corrected at higher orders in the heterotic string loop or $\alpha'$
expansions. One might speculate then that these corrections could
smooth out the singular membrane core without a source term
in the type $IIA$ theory, which would then remain a theory of only
fundamental strings by returning us to case (ii) above.
If this scenario was realized, it would ruin the exchange
of fundamental and solitonic strings between these two theories, as
well. In any event, we find this an unlikely possibility.}.

It would be of interest to determine the world-volume action
describing their dynamics of these membranes. The construction of this action
would require an examination of the zero-modes for these solutions
\strom. Since
the membranes only break half of the spacetime supersymmetries,
we know that the world-volume action will be supersymmetric.
However, a simple counting of bosonic
translational and fermionic supersymmetric degrees of
freedom \refs{\ccount,\prep}\ indicates that these membranes
will not have a conventional $\kappa$-symmetric world-volume
action \convent. One can also consider the description of these
membranes in the context of the ten-dimensional type $IIA$ theory
where they become six-branes. We were guided
by supersymmetry to choose the gauge field as one of the four vectors
in the supergravity multiplet. From the point of
view of the ten-dimensional type $IIA$ theory, three of these
fields are associated with three-form potentials which are
anti-self-dual harmonic forms on the internal $K3$ space, while
the last is a linear combination of the fundamental
vector and the dual of the spacetime three-form
potential\foot{In passing we add the following observations:
Within the heterotic string theory, the four vectors are
indistinguishable being interchanged by a discrete
$O(20,4;{\bf Z})$ transformation. This
permutation symmetry is obvious from the point of view
of their embedding in the ten-dimensional heterotic string
theory. The type $IIA$ theory inherits this permutation symmetry
since the equivalence of the two string theories dictates
that $O(20,4;{\bf Z})$ remains an exact symmetry.
However in the latter case, $O(20,4;{\bf Z})$ is permuting
fields which know about the K3 space with others that do not.}.
The membranes corresponding to the
first three fields are then ``up-lifted'' to
anisotropic six-branes with magnetic four-form charge --
note that the antiselfdual form on K3 are not localized \dpage.
The last membrane is raised to an anisotropic
six-brane carrying both conventional magnetic charge, and electric
four-form charge. Again these branes in the ten-dimensional theory
will not have a conventional Green-Schwarz world-volume action
\refs{\ccount,\prep}. Determining the correct world-volume action
would also provide an important consistency check for our analysis.
Given this action,
one could verify that the membrane solutions are also singular
for the membrane's sigma-model metric, \ie
from the point of view of membrane test-probes
-- note that given the previous discussion, we know this metric
is {\it not} that determined by the usual scaling arguments \prep.
Similarly the fundamental type $IIA$ string solution should be
singular from the membrane viewpoint.
Both of these results are necessary for a consistent
interpretation of the membranes as fundamental objects.

Given that
not much is known about the quantization of
$p$-branes for $p\ge2$, we may still consider various
scenarios by which these fundamental membranes (and possibly other
objects) would be incorporated in the type $IIA$ theory.
The complete type $IIA$ theory might be
described by one of (at least) three alternatives:

{\it An egalitarian theory of branes:}\/--
In this the simplest alternative, the full type $IIA$ theory
is a theory which contains (at least) two distinct fundamental
objects, strings and membranes. First quantization would be
separately applied for each brane with its distinct world-volume
action. A second step would be to incorporate interactions between
the different branes in this first quantized framework.
Presumably in this theory, the membranes would not contribute
to the massless spectrum at a generic point in the
(known) vacuum moduli space,
since the latter spectrum is fully accounted for by type $IIA$ strings.
In this case, the membranes would play no role
in the low energy physics, but would be important for a consistent
definition of
the theory at the level of massive modes and through nonperturbative
effects \becker. Such a egalitarian description the type $IIA$ theory
was advocated in
\towns, where in fact on the basis of $U$-duality
the democracy was extended to all $p$-branes appearing in the
low energy theory.

{\it A theory of only higher branes:}\/--
In this second scenario, the true type $IIA$ theory would actually
be a theory of only membranes (or some higher $p$-branes).
The fundamental strings would then be ``string-like''
excitations of the membrane. A similar scenario has already
been conjectured
for the heterotic string. There, certain
supersymmetric electrically charged extremal black holes
appear as singular point-like objects, but may be
identified with states in the fundamental string
spectrum \refs{\bhep,\peet}. In order for this alternative to be
consistent in the present case, the membranes must also be
able to act as sources
for the Kalb-Ramond fields that are associated with the
fundamental type $IIA$ string.  This requirement could
be confirmed by
examining the zero-mode structure of these solutions.
Further, a much more stringent constraint is that
consistently quantizing the
fundamental membranes must reproduce precisely
the same massless spectrum as the type $IIA$ string in this $K3$
context. A higher brane
description of the type $IIA$ theory
was advocated in
\town\ with the suggestion that the correct
fundamental theory was an eleven-dimensional supermembrane theory.

{\it Something else:}\/--
On this alternative, of course, we have the least to say.
However we note that past efforts at quantizing higher $p$-branes have
met with
no success. Further even if a free-first quantized theory
was constructed, the introduction of interactions for higher
$p$-branes would remain a significant challenge. These technical
obstructions lend favor to the opinion that only one-branes or strings
should be treated as fundamental. The present analysis,
which indicates that the type $IIA$ theory must incorporate
fundamental membranes, may then be an indication that the
correct fundamental description of the theory is simply not one based
on the first quantization of extended objects.

%\medskip

We
expect that other solitonic $p$-brane solutions
arising in various string theories will
also become
fundamental in the dual strong coupling theories.
Indeed, another example
is the solitonic fivebrane in $SO(32)$ heterotic
string theory, which seems to satisfy all of the necessary
criteria of spacetime
supersymmetry and no core singularity
as well as becoming
singular for the type I strings \heti, which is conjectured to
be the dual strong coupling theory in ten dimensions.

In  these examples, it is the strong
coupling heterotic string that is replaced by a weak coupling
theory that includes fundamental $p$-branes.
An interesting question
is then if the heterotic string theory stands as a complete
perturbative theory of only fundamental strings\foot{We
note that  a complementary analysis
\towns\ seems to indicate that the answer is affirmative.}.
Clearly further investigations of $p$-branes and
searches for new solutions are required to determine
the existence of $p$-brane solitons in
\eg the weakly coupled $IIA$ theory which become fundamental in
the context of heterotic strings.

However, combining
the evidence connecting
all of the critical superstring
theories in diverse dimensions and coupling regimes \ed,
the non--perturbative results in which  $p$-brane solitons
and instantons are indispensable \refs{\becker,\cone},
and
our present results indicating the inclusion of fundamental
membranes in
the type $IIA$ theory, it is perhaps
not premature to respond to the question ``Is string
theory a theory of strings?'' with the intriguing answer: No!

\vskip2truecm

\noindent
{\bf Note Added:}

After this paper was submitted for publication,
we computed how the mass per unit area of the membrane solutions
scale with the string
coupling constants of each theory, and found the following result:
$$
\mu\sim{1\over(\alpha'_{het})^{3/2}\lambda_{het}^2}
\sim{1\over(\alpha'_{II})^{3/2}\lambda_{II}}\ \ ,
$$
where $\mu$ is the ADM mass per unit area while $\lambda$ and $\alpha'$
are the coupling constants and inverse string tensions for the respective
theories indicated by the subscript. (The compactification volume is
measured in units of the corresponding $\alpha'$.) The essential point
to understanding the consistency of these results is recognizing that
varying $\lambda_{het}$ with fixed $\alpha'_{het}$ is not the same
as varying $\lambda_{II}$ with fixed $\alpha'_{II}$
-- \ie $\alpha'_{het}\lambda_{het}=\alpha'_{II}\lambda_{II}$.
(This result is also
consistent with the D--brane picture presented  in the recent paper of
Polchinski and Witten\ref\joeed{J. Polchinski and E. Witten, ``Evidence
for Heterotic --- Type I String Duality'', preprint [hep-th/9510169].}.)
Although from the mass scaling in the type $IIA$ theory it may
appear that the solution
is solitonic instead of  fundamental, we note that
a source term is still required for the membrane to be a
solution of the equations, suggesting that the theory still needs to be
supplemented by a new membrane--like object.
Further  discussion of
these results will appear in a later publication in which we will examine
the membranes in more detail.

\vskip1truecm

\noindent
{\bf Acknowledgments:}

\noindent
We would like to acknowledge illuminating discussions with
and insightful suggestions from Shyamoli Chaudhuri,
Petr Horava, Tom\'as Ort\'{\i}n,
Eric Sharpe, Eva Silverstein,
Andy Strominger and Arkady Tseytlin. We are grateful to Petr Horava
and Vipul Periwal for  comments on an earlier version of this manuscript.
We would also like to thank Arkady Tseytlin for bringing the
reference \tseyt\ to our attention.
This research was supported by NSERC of Canada, Fonds FCAR
du Qu\'ebec. CVJ would also like to thank the
McGill Physics Department for hospitality while this
research was begun; some of this work was done while CVJ was
at the Princeton University Physics department;
CVJ was supported in part by the National Science Foundation under
Grant No. PHY94-07194. RRK would like to thank the
Pennsylvania State University and the Institute for
Advanced Study for their hospitality and where
this work was completed.

\vfil\eject
\listrefs
\bye